\documentclass[a4paper,aps,intlimits,twocolumn,pre,10pt]{revtex4-1}
\usepackage{amsmath,amsthm,amssymb,amsthm,dsfont,mathrsfs}
\usepackage{mathtools,centernot}
\usepackage{tikz}
\usepackage[english]{babel}
\pdfoutput=1
\usepackage{hyperref}
\hypersetup{pdfauthor={Sicuro Rapcan Tsallis},pdftitle={NIFPE},%
            colorlinks, linktocpage=true, pdfstartpage=3, pdfstartview=FitV,%
    breaklinks=true, pdfpagemode=UseNone, pageanchor=true, pdfpagemode=UseOutlines,%
    plainpages=false, bookmarksnumbered, bookmarksopen=true, bookmarksopenlevel=1,%
    hypertexnames=true, pdfhighlight=/O,%
    urlcolor=orange, linkcolor=blue, citecolor=red, 
        }

\newcommand{\R}{\mathds{R}}

\newcommand{\N}{\mathds{N}}

\newcommand{\mE}{\mathcal{E}}
\DeclareMathOperator{\dd}{d}
\DeclareMathOperator{\e}{e}

\usepackage[T1]{fontenc}
\usepackage{lmodern}

\begin{document}
\title{Nonlinear inhomogeneous Fokker--Planck equations: Entropy and free-energy time evolution}
\author{Gabriele Sicuro}\email{sicuro@cbpf.br}
\affiliation{Centro Brasileiro de Pesquisas F\'isicas, Rua Dr. Xavier Sigaud, 150, 22290-180, Rio de Janeiro, Brazil}
\author{Peter Rap\v can}\email{rapcan@cbpf.br}
\affiliation{Centro Brasileiro de Pesquisas F\'isicas, Rua Dr. Xavier Sigaud, 150, 22290-180, Rio de Janeiro, Brazil}
\author{Constantino Tsallis}\email{tsallis@cbpf.br}
\affiliation{Centro Brasileiro de Pesquisas F\'isicas, and National Institute of Science and Technology for Complex Systems, Rua Dr. Xavier Sigaud, 150, 22290-180, Rio de Janeiro, Brazil}\affiliation{Santa Fe Institute, 1399 Hyde Park Road, Santa Fe, New Mexico, 87501 USA}
\date{\today}
\begin{abstract}We extend a recently introduced free-energy formalism for homogeneous Fokker--Planck equations to a wide, and physically appealing, class of inhomogeneous nonlinear Fokker--Planck equations. In our approach, the free-energy functional is expressed in terms of an entropic functional and an auxiliary potential, both derived from the coefficients of the equation. With reference to the introduced entropic functional, we discuss the entropy production in a relaxation process towards equilibrium. The properties of the stationary solutions of the considered Fokker--Planck equations are also discussed. \end{abstract}
\maketitle

\section{Introduction}
\label{intro}
Since the seminal work of \textcite{Einstein1905} on the Brownian motion, linear Fokker-Planck equations (FPEs) \cite{risken1996} have played a central role in the study of normal diffusion processes and in the investigation of nonequilibrium in general. It is well known, however, that many physical phenomena are associated to an anomalous diffusive behavior, that cannot be properly described by a linear FPE. For this reason, nonlinear FPEs \cite{Frank2005,Chavanis2008}, alongside with fractional linear FPEs \cite{Metzler1999,Barkai2000}, have become natural candidates for modeling anomalous diffusion processes. Models based on nonlinear FPEs are indeed able to reproduce the experimentally observed dispersion laws. In the last decades, nonlinear FPEs have been put in relation with generalized thermostatistics \cite{Plastino1995} and successfully describe diffusion in porous media \cite{aronson1986}, {stellar dynamics and turbulence \cite{Chavanis2003}}, or surface dynamics \cite{Spohn1993}. Similarly, fluctuations in granular media can be properly treated by means of nonlinear FPEs \cite{Tsallis1996}, as recently experimentally verified by \textcite{Combe2015}. In Ref.~\cite{Borland2002} a nonlinear FPE was adopted to model the evolution of stock price returns, finding a remarkable agreement with the market data. 

The reconstruction of the microscopical dynamics corresponding to a given nonlinear FPE is, however, a nontrivial task. In this regard, \textcite{Borland1998} proposed a phenomenological model, in which the evolution at the microscopic level can be simulated to successfully reproduce the macroscopic quantities: the equations of motion for the microscopic components, however, depend on the solution of the nonlinear FPE itself. Macroscopic and microscopic evolution are therefore coupled, suggesting that the model can be used as a heuristic description only, as stressed by the author herself.

The nontrivial relation between macroscopic and microscopic evolution in nonlinear FPEs may be relevant in the study of thermodynamics. {To be more precise, let us recall that a diffusion process in $d$ dimensions can be studied in the $2d$-dimensional one-particle phase space, considering a particle distribution density $f(\mathbf r,\mathbf v,t)$ around the space position $\mathbf r$ and the velocity value $\mathbf v$ at time $t$. The evolution of $f$ is described by the so-called Klein--Kramers equation (KKE, or FPE in the one-particle phase space) \cite{Kramers1940}, having the form
\begin{equation}\label{kramers}
\frac{\dd f}{\dd t}=\left[\frac{\partial}{\partial t}+\mathbf v\cdot\nabla_\mathbf{r}+\frac{1}{m}\mathbf F(\mathbf r)\cdot\nabla_\mathbf{v}\right]f=K[f],
\end{equation}
where $\mathbf F(\mathbf r)$ is an external force acting on the particle of mass $m$, and $K[f]$ is a functional of $f$ determined by the underlying kinetics. The Boltzmann equation is a particular case of Eq.~\eqref{kramers}. The possible emergence of nonlinearity in the KKE is due to the structure of the functional $K[f]$. \textcite{Kaniadakis2001} proposed a very general kinetic interaction principle that is able to unify many relevant particular cases in one single picture. He also associated to $f$ an entropic functional $\mathcal S(f)$, satisfying the $H$ theorem such that 
\begin{equation}
\frac{\dd f}{\dd t}+\nabla_\mathbf{v}\cdot\left[D(\mathbf v)\gamma(f)\nabla_\mathbf v\frac{\delta \mathcal S(f)}{\delta f}\right]=0,
\end{equation}
where $D(\mathbf v)$ is a velocity-dependent diffusion coefficient and $\gamma$ is a function of $f$ only, related to the kinetics $K$. The relation between a KKE and the corresponding equations for $\rho_r(\mathbf r,t)=\int f(\mathbf r,\mathbf v,t)\dd^dv$ and $\rho_{v}(\mathbf v,t)\coloneqq \int f(\mathbf r,\mathbf v,t)\dd^dr$ (usually called Smoluchowski equation, or SE, and FPE, respectively) is however not trivial \cite{Wilemski1976}. For example, for a Brownian particle in a fluid, we have \cite{risken1996,Kaniadakis2001}
\begin{equation}\label{kbm}
K[f]=\lambda\nabla_\mathbf{v}\cdot\left(\mathbf v f+\lambda D\nabla_\mathbf{v}f\right).
\end{equation}
In the expression above, $\lambda$ is a friction coefficient and $D>0$ is a diffusion constant. The SE corresponding to the KKE obtained using Eq.~\eqref{kbm} is typically written as
\begin{equation}\label{smolu}
\frac{\partial \rho_r}{\partial t}+\nabla_\mathbf{r}\cdot\left(\frac{\rho_r\mathbf F}{{\lambda m}}+D\nabla_\mathbf r\rho_r\right)=0,
\end{equation}
to be associated with a proper initial condition $\rho_r(\mathbf r,0)=\int f(\mathbf r,\mathbf v,0)\dd^dv$. This equation, however, is obtained assuming $\lambda \gg 1$. It is expected that, for a generic value of $\lambda$, the evolution of $\rho_r$ should depend on $f(\mathbf r,\mathbf v,0)$ and not on $\rho_r(\mathbf r,0)$ only, and that, therefore, Eq.~\eqref{smolu} must be corrected. A first study of the corrections to Eq.~\eqref{smolu} was performed by \textcite{Wilemski1976}, and later by \textcite{Chaturvedi1979,Miguel1980} in the case of a position dependent external force in the one dimensional case. In Refs.~\cite{Ferrari2008,Ferrari2010} the exact SE corresponding to a given KKE was obtained, in which the coefficients of the SE depend on the coefficients of the KKE and, moreover, on its initial condition $f(\mathbf r,\mathbf v,0)$.} 

{With these considerations in mind, it is clear that a thermodynamical approach based on the coefficients of a FPE or a SE, even in the linear case, is not equivalent (in general) to a direct investigation on the corresponding KKE, and that a SE or a FPE are usually obtained under certain hypotheses and subject to corrections. In the limits of validity of a SE or a FPE, however, a thermodynamical functional approach remains of great interest and can give useful information about both the relaxation process and fluctuations.} For example, \textcite{Bertini2015} recently introduced a remarkable set of ideas, now called macroscopic fluctuation theory, for a nonequilibrium thermodynamics of systems described by a general evolution law of the type of a nonlinear SE \cite{Bertini2012}, avoiding reference to the microscopical details.

A different approach to the thermodynamics of nonlinear FPEs \footnote{{In the following we will not distinguish between FPE and SE, due to the fact that they have a similar mathematical structure. We will simply say that the considered equation is a FPE.}} has been independently proposed by \textcite{schwammle2007,schwammle2007b}. {Their formalism is analogous to the one introduced in Ref.~\cite{Kaniadakis2001} for KKEs and it was inspired by the contribution of \textcite{Plastino1995}.} They considered a nonlinear FPE for a probability density $\rho$ in $(1+1)$ dimensions, in the form
\begin{equation}
\begin{cases}
\partial_t\rho(x,t)+\partial_x j(x,\rho)=0,\\
j(x,\rho)=\chi\left[\rho(x,t)\right]E(x)-D\omega\left[\rho(x,t)\right]\partial_x\rho(x,t),\\
\lim_{x\to \pm\infty}j(x,\rho)=0.
\end{cases}
\end{equation} 
Here $\chi(\rho)$ and $\omega(\rho)$ are positive quantities depending on the density $\rho$ only, $D$ is a positive diffusion constant, and $E(x)$ is an external field. They showed that it is possible to construct a free-energy functional $F(\rho)$ which is consistent with thermodynamics and which satisfies $\partial_t F(\rho)\leq 0$. The expression for $F(\rho)$ is given explicitly in terms of the coefficients of the equation. Following this approach, nonlinear FPEs are typically (but not always) associated to entropic functionals that are different from the Boltzmann--Gibbs entropy, obtained in the linear case. The stationary distribution of a given nonlinear FPE coincides with the one that maximizes the corresponding entropy with an appropriate energy constraint. Due to the fact that nonlinear FPEs appear in the study of vortex diffusion in superconductors, and inspired by the formalism above, \textcite{andrade2010} claimed that a nonextensive thermostatistics, different from the Boltzmann--Gibbs one, is necessary for the study of the overdamped motion of interacting vortices at zero temperature \cite{Ribeiro2012,Vieira2016}. 

In the present paper we show that the approach of Schw\"ammle, Curado and Nobre can be generalized to \textit{inhomogeneous} nonlinear FPEs, i.e., to nonlinear FPEs having a diffusion coefficient $D$ depending on $x$, $D=D(x)$. Nonlinear inhomogeneous FPEs describe, for example, anomalous diffusion processes in which, in addition to the nonlinearity, a local inhomogeneity of the medium is present, inducing a space dependent friction coefficient \cite{Lau2007}. {Position dependent, or velocity dependent, diffusion coefficients are not uncommon in the literature. For example, in Ref.~\cite{Kaniadakis2001} a velocity dependent diffusion coefficient appears in the study of a generalized kinetics, developed in the proper one-particle phase space. We will focus, however, on the Fokker--Planck picture.} In Sec.~\ref{sec:freenergy} we introduce, in particular, a functional $\Phi$ that has (locally) the structure of a free energy rescaled by $D(x)$. In the linear homogeneous case, assuming the Einstein--Smoluchowski relation $D\propto T$ between the diffusion coefficient and the temperature of the bath, our rescaled functional reduces to $\Phi\propto T^{-1}F$, where $F$ is the usual free energy. We also study the time evolution of the entropy in a relaxation process towards equilibrium. In Sec.~\ref{sec:stat} we discuss the relation between $\Phi$ and the stationary solution of the considered nonlinear FPE, and a possible definition of a temperature-like quantity in this context. In Sec.~\ref{sec:poro}, we study, as a particular example, a FPE for diffusion processes in inhomogeneous porous media. Finally, in Sec.~\ref{sec:conclusioni}, we give our conclusions.

\section{Modified free-energy functional}\label{sec:freenergy}
A general nonlinear Fokker--Planck equation in $(1+1)$ dimensions describes the evolution with time of a probability density function $\rho(x,t)$ defined on the open interval $\Xi\coloneqq(x_-,x_+)$ of the real line. We admit $\Xi\equiv \R$ as a particular case. The equation has the general form of a probability conservation law
\begin{subequations}\label{fp}
\begin{equation}\label{consrho}
\frac{\partial \rho(x,t)}{\partial t}+\frac{\partial j(x,\rho)}{\partial x}=0.
\end{equation}
The current of probability $j$ introduced above has the structure \cite{Bertini2015,Eyink1990,Kipnis2013}
\begin{equation}\label{J}
j(x,\rho)\coloneqq E(x)\chi[\rho(x,t)]-D(x)\omega[\rho(x,t)]\partial_x\rho(x,t),
\end{equation}\end{subequations}
where $\chi(\rho)>0$ is the \textit{mobility} and $E(x)$ is a \textit{drift coefficient} related to the presence of an external potential $V(x)$,
\begin{equation}\label{phi}
E(x)=-\frac{\dd V(x)}{\dd x}.
\end{equation}
In the present paper, we assume that the \textit{diffusion coefficient} $\mathcal D(x,\rho)\coloneqq D(x)\omega(\rho)$ in Eq.~\eqref{J} is in this specific factorized form; we will also assume that both the factors are strictly positive, i.e., $D(x)>0$ and  $\omega(\rho)>0$, for (almost) all values of their arguments. 
Equation \eqref{J} is typically obtained through a set of approximations from a microscopical model and assuming a linear response to the action of the external field. The type of approximations strongly depend on the considered model and on the assumptions about the underlying dynamics. Moreover, in the one-dimensional case the conservation of probability implies some additional constraints on $j$, namely, the fact that the current has the same value on the boundary for all values of $t$. We will consider \textit{reflecting boundary conditions} \cite{gardiner2004} for the probability current $j$, i.e., our problem has the form
\begin{equation}
\begin{cases}
\partial_t\rho(x,t)+\partial_x j(x,\rho)=0,\\
j(x,\rho)=\chi\left[\rho(x,t)\right]E(x)-D(x)\omega\left[\rho(x,t)\right]\partial_x\rho(x,t),\\
\lim_{x\to x_\pm}j(x,\rho)=0.
\end{cases}\label{R}
\end{equation} 
Reflecting boundary conditions imply that the stationary state $\varrho(x)$ has $j(x,\varrho)=0$ on the entire domain, i.e., the stationary solution is the equilibrium solution, and nonequilibrium stationary solutions are not allowed \footnote{In the present paper, we say that a solution $\varrho$ is \textit{stationary} if $j(x,\varrho)=\text{constant}$ on the domain $\Xi$. In particular, a stationary solution is an equilibrium solution if $j(x,\varrho)=0$ on $\Xi$.}. 

\textcite{schwammle2007,schwammle2007b} observed that a FPE in the form in Eq.~\eqref{fp} can be associated to a trace-form free-energy-like functional decreasing in time. In the original paper, $D$ was supposed to be a constant. Inspired by their result, we will show that a similar functional can be obtained in the inhomogeneous case. In particular, we search for a functional in the form
\begin{equation}\label{free}
\Phi(\rho)\coloneqq\int_{\Xi}\phi\left[x,\rho(x,t)\right]\dd x\coloneqq \bar U(\rho)-S(\rho),
\end{equation}
such that the following inequality holds,
\begin{equation}
\frac{\dd  \Phi}{\dd t}\leq 0, \quad t\geq 0.\label{Hth}
\end{equation}
Observe that $\Phi$ has the structure of a free-energy rescaled by the temperature. The first term 
\begin{equation}
\bar U(\rho)\coloneqq\int_{\Xi}\bar{V}(x)\rho(x,t)\dd x
\end{equation}
corresponds to an ``energy contribution'' expressed in terms of an \textit{auxiliar} potential $\bar{V}(x)$, in general different from $V(x)$. The second term corresponds to an entropic contribution 
\begin{equation}\label{entropia}
S(\rho)\coloneqq\int_{\Xi}s[\rho(x,t)]\dd x,\quad s(0)=s(1)=0.
\end{equation}
Both the form of $\bar{V}(x)$ and of $s(\rho)$ can be determined by imposing the condition in Eq.~\eqref{Hth}. This result is sometimes called the ``$H$ theorem'' in the literature \cite{frank2001,schwammle2007,casas2016}. 
In particular, the Boltzmann--Gibbs entropy will be recovered for linear FPEs, while other commonly used generalized entropies are naturally associated to a wide class of nonlinear FPEs \cite{schwammle2007,Ribeiro2013,Ribeiro2014}. Differentiating Eq.~\eqref{free} we have
\begin{multline}\label{dFdt}
\frac{\dd  \Phi}{\dd t}=\int_{\Xi}\frac{\partial \rho}{\partial t}\left.\frac{\partial \phi(x,y)}{\partial y}\right|_{y=\rho}\dd x\\
=-\int_{\Xi} \left[-E(x)\chi(\rho)+D(x)\omega(\rho)\frac{\partial \rho}{\partial x}\right]\frac{\partial}{\partial x}\left.\frac{\partial \phi(x,y)}{\partial y}\right|_{y=\rho}\dd x\\
=-\int_{\Xi}\dd x\ D(x)\chi(\rho)\left[-\frac{E(x)}{D(x)}+\frac{\omega(\rho)}{\chi(\rho)}\frac{\partial \rho}{\partial x}\right]\\
\times \left[\frac{\dd\bar{V}(x)}{\dd x}-\left.\frac{\dd^2 s(y)}{\dd y^2}\right|_{y=\rho}\frac{\partial \rho}{\partial x}\right].
\end{multline}
We have omitted the explicit dependency on $x$ and $t$ in $\rho$ for simplicity of notation. Comparing the previous result with the condition in Eq.~\eqref{Hth}, we have that the inequality is always satisfied if
\begin{subequations}\label{EqF}
\begin{align}
\frac{\dd\bar{V}(x)}{\dd x}&=-\frac{E(x)}{D(x)}=\frac{1}{D(x)}\frac{\dd V(x)}{\dd x},\label{EqF1}\\
-\frac{\dd^2s(\rho)}{\dd \rho^2}&=\frac{\omega(\rho)}{\chi(\rho)},\quad s(0)=s(1)=0.
\end{align}
\end{subequations}
Similar equations have been obtained by \textcite{schwammle2007,schwammle2007b} in the case of a homogeneous Fokker--Planck equation, i.e., $D(x)\equiv D=\text{constant}$. In particular, we recover the same expression for the entropy, while the potential is replaced by a more general ``effective'' potential that reduces to the usual one in the homogeneous case.  Indeed, in the case of a constant friction coefficient, the natural identification $\bar{V}(x)\equiv \beta V(x)$ holds, where $\beta=D^{-1}$. Being $D(x)>0$, if $V(x)$ is bounded from below, then $\bar{V}(x)$ is bounded from below as well. This implies that the functional $\Phi$ is bounded from below. We stress again that the expression for $\Phi$ and the inequality in Eq.~\eqref{Hth} has been obtained assuming zero current on the boundary of the domain. 

\subsection{Properties of the functional $\Phi$}
Let us now discuss some properties of the function $\Phi$. As in the homogeneous case, $s(\rho)$ is strictly concave, and the rescaled free-energy $\Phi$ is strictly convex with respect to $\rho$,
\begin{equation}
\frac{\delta^2 \Phi}{\delta \rho^2}=-\left.\frac{\dd^2 s(y)}{\dd y^2}\right|_{y= \rho(x,t)}=\frac{\omega(\rho)}{\chi(\rho)}> 0.
\end{equation}
The relation above can be written as $\omega(\rho)=\chi(\rho)\partial_\rho^2\phi(x,\rho)$. Moreover, Eq.~\eqref{fp} becomes
\begin{multline}\label{fpbis}
\frac{\partial \rho}{\partial t}=\frac{\partial}{\partial x}\left(\chi(\rho) D(x)\frac{\partial}{\partial x}\frac{\delta\Phi}{\delta\rho}\right),\\ j(x,\rho)=-D(x)\chi(\rho) \frac{\partial}{\partial x}\frac{\delta \Phi}{\delta\rho}.
\end{multline}
In a relaxation process towards the equilibrium distribution $\varrho(x)$, using the boundary condition in Eq.~\eqref{R}, the time derivative of the entropy $S(\rho)$ is
\begin{equation}\label{St}
\frac{\dd S(\rho)}{\dd t}
=\int_\Xi\frac{j^2(x,\rho)}{\chi(\rho) D(x)}\dd x- \int_\Xi \frac{j(x,\rho)E(x)}{D(x)}\dd x.
\end{equation}
Equation \eqref{St} is, up to a global positive factor, a generalization of a corresponding expression obtained by \textcite{casas2012} for the case $D(x)=D=\text{constant}$. In particular, the first integral can be identified with the entropy production contribution, and it is always positive. Both terms in the equation approach zero as $\rho(x,t)\to \varrho(x)$. Finally, introducing the average energy
\begin{equation}
U(\rho)\coloneqq \int_\Xi V(x)\rho(x,t)\dd x,
\end{equation}
the energy dissipation rate is given by
\begin{equation}\label{Udiss}
\dot U=-\int_\Xi E(x)j(x,\rho)\dd x.
\end{equation}

In the case of a homogeneous medium, $D(x)\equiv D$, we can define the free energy density and the free energy as 
\begin{equation}\label{freevera}
f(x,\rho)\coloneqq D\phi(x,\rho),\quad F(\rho)\coloneqq\int_{\Xi} f(x,\rho)\dd x,
\end{equation}
respectively. The free energy density satisfies the relation
\begin{equation}
\mathcal D(\rho)\coloneqq D\omega(\rho)=\chi(\rho)\partial_\rho^2 f(x,\rho),\label{Einstein}
\end{equation}
that has the structure of a local Einstein fluctuation--dissipation relation. Moreover, the FPE can be written as
\begin{equation}
\frac{\partial \rho}{\partial t}=\frac{\partial}{\partial x}\left(\chi(\rho) \frac{\partial}{\partial x}\frac{\delta F}{\delta\rho}\right),\quad j(x,\rho)=-\chi(\rho) \frac{\partial}{\partial x}\frac{\delta F}{\delta\rho}.
\end{equation}
In the homogeneous case, Eq.~\eqref{St} has a more clear interpretation. Indeed, in an electromagnetic analogy, we can think of $j$ as a current of charges in an external electric field $E$, flowing in a medium whose resistance is given by $\chi$. Therefore, the first term plays the role of a dissipation power contribution, whilst the second term is related to the rate of energy exchange between the external field and the charges, and it indeed coincides with Eq.~\eqref{Udiss} up to a constant multiplicative factor.

\section{Equilibrium solution on the real line}\label{sec:stat}
Let us assume now that our domain is the real line, i.e., $\Xi\equiv\R$. Adopting reflecting boundary conditions, it is immediately seen that the equation for the stationary solution of Eq.~\eqref{fp} is the equation for the equilibrium state $\varrho$, i.e.
\begin{equation}\label{eqstat}
j(x,\varrho)=0\Leftrightarrow
\frac{\dd \varrho(x)}{\dd x}=\frac{E(x)\chi(\varrho)}{D(x)\omega(\varrho)}.
\end{equation}
On the other hand, because of the required integrability of $\varrho$, we ask
\begin{equation}
\lim_{x\to\pm\infty}\varrho(x)=\lim_{x\to\pm\infty}\frac{\dd \varrho(x)}{\dd x}=0.
\end{equation}

To prove that the limit distribution, for any initial condition, is uniquely identified, and coincides with $\varrho$, we first follow the arguments of \textcite{frank2001} for the homogeneous case, with proper modifications to adapt them to our case. From Eq.~\eqref{eqstat}, we have
\begin{equation}\label{eqstat2}
\frac{E(x)}{D(x)}=-\frac{\dd \bar{V}(x)}{\dd x}=\frac{\omega(\varrho)}{\chi(\varrho)}\frac{\dd \varrho(x)}{\dd x}=-\frac{\dd}{\dd x}\left[\left.\frac{\dd s(z)}{\dd z}\right|_{z=\varrho}\right].
\end{equation}
Let us now introduce the function
\begin{equation}\label{sigma}
\exp_s(x)\coloneqq \left[\frac{\dd s}{\dd x}\right]^{-1}(-x),
\end{equation}
inverse of the function $\partial_p s(p)$ evaluated in $-x$. The function in Eq.~\eqref{sigma} exists, being  $\partial_\rho^2 s(\rho)<0$ strictly, i.e., $\partial_\rho s(\rho)$ is strictly decreasing for $\rho>0$. Denoting by 
\begin{equation}
x_0\coloneqq-\lim_{\rho\to 0}\partial_\rho s(\rho)<0,
\end{equation} 
the function $\exp_s(x)$ is positive and strictly increasing in $(x_0,+\infty)$ (here $x_0$ can be also not finite) and we impose that it is identically zero in $(-\infty,x_0)$, being $\lim_{x\to x_0^+}\exp_s(x)=0$. For future convenience, we define also
\begin{equation}
\log_s(x)=\exp_s^{-1}(x),\quad \log_s(x)\colon\R^+_0\to (x_0,+\infty),
\end{equation}
inverse function of $\exp_s(x)$. For a linear FPE, $s(p)=-p\ln p$, then $\exp_s(x)=\e^{x-1}$ and $\log_s(x)=\ln x+1$. By means of the introduced function, we can write, for some $c$ to be determined,
\begin{equation}\label{statsigma}
\varrho(x)=\exp_s\left[c-\bar{V}(x)\right].
\end{equation}
In the previous expression, we have supposed that the arbitrary additive constant in $\bar{V}$ is somehow fixed. It is important to prove that the normalization constant $c$ in Eq.~\eqref{statsigma} exists, and that it is uniquely identified. For this purpose, observe that the function
\begin{equation}
h(y)\coloneqq\int_{-\infty}^{+\infty}\exp_s\left[y-\bar{V}(x)\right]\dd x
\end{equation}
is strictly monotonically increasing, being
\begin{equation}
h'(y)=-\int_{-\infty}^{+\infty}\frac{\dd x}{s''\left\{\exp_s\left[y-\bar{V}(x)\right]\right\}}>0.
\end{equation}
Moreover, we have that $\lim_{y\to -\infty}h(y)=0$ and
\begin{multline}
h(y)\geq \int_{-\infty}^{+\infty}\theta\left(y-1+\bar{V}(x)\right)\exp_s\left[y-\bar{V}(x)\right]\dd x\\
\geq \exp_s(1)\int_{-\infty}^{+\infty}\theta\left(y-1+\bar{V}(x)\right)\dd x\xrightarrow{y\to\infty}+\infty.
\end{multline}
It follows that the normalization constant exists and it is unique. The constant $c$, uniquely identified, satisfies the identity $c\equiv \bar{V}(x) +\log_s\left[\varrho(x)\right]$. Using the fact that $\bar{V}$ is defined up to an additive constant, we can absorb $c$ in the auxiliar potential, in such a way that
\begin{equation}
\bar{V}(x)=-\log_s\left[\varrho(x)\right],
\end{equation}
This gives $\delta_\rho \Phi|_{\rho=\varrho}=0$. Moreover, the convexity of $\Phi$ implies that $\varrho$ is a minimum for $\Phi$. On the other hand, from Eq.~\eqref{dFdt} we have that
\begin{equation}\label{HthSta}
\left.\frac{\dd \Phi}{\dd t}\right|_{\rho=\varrho}=0.
\end{equation}
The uniqueness of the stationary solution in the hypotheses specified, together with the result in Eq.~\eqref{Hth}, guarantees that the limit distribution is the stationary distribution.

This result can be proven in a different, and more explicit, way, without invoking Eq.~\eqref{Hth}. Let us suppose that the entropy density $s(\rho)$ is known from the coefficients in Eq.~\eqref{fp}. From the functional $S(\rho)$, and following again the approach of \textcite{frank2001}, we can construct the following generalized divergence between two probability distribution densities $\rho_1=\rho_1(x,t)$ and $\rho_2=\rho_2(x,t)$,
\begin{multline}
\Delta_s(\rho_1\|\rho_2)\coloneqq S(\rho_2)-S(\rho_1)+\\
+\int_{-\infty}^{+\infty}\left[\left(\rho_1(x,t)-\rho_2(x,t)\right)\left.\frac{\partial s(z)}{\partial z}\right|_{z=\rho_2}\right]\dd x.
\end{multline}
The quantity $\Delta_s$ is always non-negative, due to the strict concavity of the function $s(\rho)$. In particular, $\Delta_s(\rho_1\|\rho_2)=0\Leftrightarrow \rho_1=\rho_2$. In the case of Boltzmann--Gibbs entropy, $s(\rho)=-\rho\ln \rho$, $\Delta_s$ is the Kullback--Leibler divergence \cite{kullback1951}
\begin{equation}
\Delta_\text{BG}(\rho_1\|\rho_2)\equiv D_\text{KL}(\rho_1\|\rho_2)\coloneqq\int_{-\infty}^{+\infty}\rho_1(x,t)\ln\frac{\rho_1(x,t)}{\rho_2(x,t)}\dd x.
\end{equation}
If we consider now the case $\rho_2\equiv \varrho(x)$, unique stationary solution of our equation, and $\rho_1=\rho(x,t)$ a solution at the time $t$ of the considered nonlinear FPE for a given initial condition, we have
\begin{equation}
\frac{\dd \Delta_s(\rho,\varrho)}{\dd t}
=-\int_{-\infty}^{\infty}D(x)\chi(\rho)\left(\left.\frac{\omega(\varrho)}{\chi(\varrho)}\frac{\partial \varrho}{\partial x}\right|_{\varrho=\rho}^{\varrho=\varrho}\right)^2\dd x\leq 0.
\end{equation}
It follows then that
\begin{equation}
\lim_{t\to\infty}\rho(x,t)=\varrho(x).
\end{equation}
\subsection{Effective temperature}
It is clear that, in the general case, a definition of a temperature-like quantity is difficult, and the meaning itself of ``temperature'' might depend on the specific considered model \cite{biro2011}. A possible definition of effective inverse temperature $\bar \beta$, however, can be given by the usual thermodynamic relation
\begin{equation}
\bar\beta\coloneqq\frac{\partial S(\varrho)}{\partial U(\varrho)},
\end{equation}
expressing with the variation of entropy respect to the average energy with reference to the equilibrium state, which is supposed to be unique. Observing that $\varrho$ is a minimum of the functional $\Phi(\rho)$, we can write the previous relation as
\begin{equation}\label{betabis}
\bar\beta=\frac{\partial\bar U(\varrho)}{\partial U(\varrho)}.
\end{equation}
In particular, if $D(x)\equiv D=\text{constant}$, then we obtain the well-known result
\begin{equation}
\bar \beta=\frac{1}{D}>0.
\end{equation}

\section{A nonlinear FPE for diffusion in inhomogeneous porous media}\label{sec:poro}
From the observations above, it is evident that the knowledge of the external potential $V(x)$ and of the entropic form $s(\rho)$ is \textsl{not} sufficient to identify the stationary distribution. On the other hand, the observation of a specific limit distribution does not allow us to infer the entropic form, unless a careful investigation of the structure of the effective potential is performed. This simple fact has been already pointed out in the study of elementary probabilistic toy models \cite{rodriguez2008,Bergeron2014,Sicuro2015}. To further exemplify it and apply our formalism, let us consider, for example, the nonlinear inhomogeneous FPE for a fluid in a porous medium. The conservation of mass imposes
\begin{equation}
\frac{\dd \rho(x,t)}{\dd t}=\frac{\partial \rho(x,t)}{\partial t}+\frac{\partial }{\partial x}\left[\dot x\rho(x,t)\right]=0
\end{equation}
for the density $\rho(x,t)$. On the other hand, in the case of diffusion in a porous medium, Darcy's law holds \cite{aronson1986},
\begin{equation}
\dot x=E(x)-\kappa(x)\frac{\partial P(x,t)}{\partial x},\quad\kappa(x)>0,
\end{equation}
where $P(x,t)$ is the pressure of the fluid at the position $x$ and at time $t$, and $E(x)=-\partial_xV(x)$ is an external force (we are working in the overdamping limit) resulting from an external potential $V(x)$. The dependence of the coefficient $\kappa(x)$ on $x$ expresses exactly the lack of homogeneity of the medium (e.g., a porosity depending on $x$). Imposing  the equation for polytropic gases
\begin{equation}
P(x,t)=\alpha \rho^{\lambda}(x,t),\quad\lambda>0,\ \alpha>0,
\end{equation}
we get the porous medium equation in the presence of an external field
\begin{equation}\label{porous}
\frac{\partial \rho(x,t)}{\partial t}=\frac{\partial}{\partial x}\left[-E(x)\rho(x,t)+D(x)\rho^{\nu-1}(x,t)\frac{\partial \rho(x,t)}{\partial x}\right],
\end{equation}
where we have defined $\nu=\lambda+1$ to uniform our notation with that adopted in Refs.~\cite{Plastino1995,Tsallis1996}, and
\begin{equation}
D(x)\coloneqq\alpha(\nu-1)\kappa(x).
\end{equation}
{Diffusion equations similar to Eq.~\eqref{porous} have been proposed, for example, in Refs.~\cite{Plastino1995,Kaniadakis2000,Kaniadakis2001} with $E(x)=-\gamma x$, $\gamma>0$, for the velocity distribution of one particle in one dimension. In particular, \textcite{Plastino1995} considered the case $D(x)\equiv D>0$. \textcite{Kaniadakis2000} assumed a time-dependent diffusion coefficient $D(t)$ that is homogeneous in space. In Ref.~\cite{Kaniadakis2000} the linear ($\nu=1$) inhomogeneous case is also considered.} Equations \eqref{EqF} take the form
\begin{subequations}
\begin{align}
\frac{\dd\bar{V}(x)}{\dd x}&=-\frac{E(x)}{D(x)},\\
\frac{\dd^2s(\rho)}{\dd \rho^2}&=-\rho^{\nu-2},\quad s(0)=s(1)=0.
\end{align}
\end{subequations}
The last equation gives, in particular,
\begin{equation}
s(\rho)=\frac{\rho-\rho^{\nu}}{\nu(\nu-1)}\equiv \frac{s_{\nu}(\rho)}{\nu},
\end{equation}
where $s_q(\rho)$ is the nonadditive entropy introduced in Refs.~\cite{havrda1967,Tsallis1988}
\begin{equation}
S_q(\rho)\coloneqq \int_{-\infty}^{+\infty}s_q[\rho(x,t)]\dd x=\frac{1-\int_{-\infty}^{+\infty}\rho^{q}(x,t)\dd x}{q-1}.
\end{equation}
Equation \eqref{eqstat} gives the stationary solution
\begin{equation}\label{stporo}
\varrho(x)= \rho_0 \e_{2-\nu}^{-\frac{\bar{V}(x)-\bar{V}(0)}{\rho_0^{\nu-1}}},
\end{equation}
where $\rho_0$ is fixed by the normalization condition. In the previous expression we have introduced the so-called \textit{$q$-exponential} function with $q\in\R$,
\begin{equation}\label{qexp}
\e_q^{x}\coloneqq\left[1+(1-q)x\right]_+^\frac{1}{1-q},\quad \text{with }[x]_+\coloneqq x\,\theta(x).
\end{equation}
The $q$-exponential is indeed related to the $\exp_{s_q}(x)$ in Eq.~\eqref{sigma}, being  
\begin{equation}
\exp_{s_q}(x)=\e_{2-q}^{\frac{x-1}{q}}.
\end{equation}
Even at fixed entropic form, we can therefore obtain a wide class of limit distributions by an appropriate choice of the argument of the $q$-exponential, and in particular of $D(x)$. Namely, to have a limit distribution $\varrho(x)$, it suffices that
\begin{equation}
  \frac{E(x)}{D (x)}=[\varrho(x)]^{\nu-2}\frac{\dd \varrho(x)}{\dd x} .
\end{equation}

As an example, let us consider a \textit{$q$-Gaussian} limit distribution
\begin{equation}
  \varrho_q(x)=\frac{\sqrt{a}}{C_{q}} \e_q^{-a x^2},\quad a>0,\quad q<3,
\end{equation}
where we have introduced the normalization constant
\begin{equation}
C_q\coloneqq\begin{cases}
\frac{{2 \sqrt{\pi} \Gamma\left(\frac{1}{1-q}\right)}}{{(3-q) \sqrt{1-q} \Gamma\left(\frac{3-q}{2(1-q)}\right)}} &\text{if } q<1,\\
\sqrt{\pi}&\text{if } q=1,\\
\frac{\sqrt{\pi} \Gamma\left(\frac{3-q}{2(q-1)}\right)}{\sqrt{q-1} \Gamma\left(\frac{1}{q-1}\right)} &\text{if }q\in(1,3).
\end{cases}
\end{equation}
The Gaussian distribution with parameter $a>0$ is recovered as a special case for $q\to 1$. It is immediately seen that the following relation between $E$ and $D$ must hold,
\begin{equation}
D(x)=-\frac{\left[1-a(1-q)x^2\right]_+}{2ax[\varrho_q(x)]^{\nu-1}}E(x).
\end{equation} 
If, for example, we consider the case of Boltzmann--Gibbs entropy ($\nu=1$), we have
\begin{equation}
D(x)=-\frac{\left[1-a(1-q)x^2\right]_+}{2ax}E(x).
\end{equation}
In particular, in the presence of an external harmonic potential,
\begin{equation}
V(x)=\frac{\gamma x^2}{2}\Rightarrow E(x)=-\gamma x,\quad \gamma>0,
\end{equation}
it is sufficient to modulate $D(x)$ as
\begin{equation}
D(x)=\gamma\frac{\left[1-a(1-q)x^2\right]_+}{2a},\quad q\geq 1,
\end{equation}
to obtain as a stationary distribution a $q$-Gaussian, despite the fact that the entropy associated to the considered equation is the Boltzmann--Gibbs entropy and the external potential is harmonic. Observe also that for $q\to 1^+$ we recover the well known result $D(x)=\text{constant}$. A particular case, in a different formalism, has been analyzed in Refs.~\cite{Kaniadakis2000,Zochil2014} for a specific choice of the coefficients $E(x)$ and $D(x)$, that indeed satisfy the relation above.
\section{Conclusions and perspectives}\label{sec:conclusioni}
In the present paper we have discussed a generalization to the inhomogeneous case of the free-energy formalism introduced by \textcite{schwammle2007} for the study of nonlinear FPEs. We have shown that a modified free-energy functional $\Phi$, defined on the space of distributions, can be introduced, in such a way that the stationary solution is a minimum for $\Phi$. This functional is explicitly expressed in terms of the coefficients of the considered FPE, and it involves an auxiliary potential and an entropic density. We have also shown that, in a relaxation process towards the stationary distribution on the real line, $\Phi$ decreases monotonically, reaching the minimum value on the stationary distribution. Some basic properties of the stationary solutions of nonlinear FPEs have been analyzed. In particular, the solutions can be expressed in terms of a generalized exponential function associated to the entropy, having the auxiliary potential appearing in $\Phi$ as argument. We have then applied our formalism to a nonlinear FPE for the macroscopic description of diffusion processes in inhomogeneous porous media.

The full understanding of the thermodynamical meaning of the discussed free-energy functionals is still an active research topic \cite{Nobre2015}. {As discussed in the Introduction, a FPE is usually obtained from a KKE that properly describes the evolution of the density in the one-particle phase space. In this sense the relation between the thermodynamical functionals introduced for a general nonlinear FPE and the ones for the corresponding KKE deserves further analysis}. Further investigations are also needed in light of the fast-developing field of stochastic thermodynamics \cite{Seifert2008,Seifert2012} and in relation to macroscopic fluctuation theory \cite{Bertini2012,Bertini2015}. The possible connections between the entropy associated to a FPE following the recipe presented here, and fluctuation theory might shed new light on foundational aspects of thermodynamics, apart from possible experimental applications. We hope to address these problems in future publications.

\section{Acknowledgments}
The authors thank Fernando D.~Nobre {and Claudio Landim} for useful discussions. The authors acknowledge the financial support of the John Templeton Foundation.

\bibliography{Biblio.bib}
\end{document}